\documentclass[reprint,aps,prx,superscriptaddress,amsmath,amssymb,floatfix,footinbib,longbibliography]{revtex4-1}
\usepackage[english]{babel}
\usepackage{graphicx}
\usepackage{xcolor}
\usepackage{bm}
\usepackage[colorlinks, linkcolor= blue, citecolor = blue, urlcolor=blue]{hyperref}
\usepackage{ulem}
\usepackage{physics}
\usepackage{wrapfig}
\usepackage[stable]{footmisc}
\usepackage{array}
\usepackage{multirow}
\usepackage{hhline}
\usepackage{pifont}
\usepackage{chemformula}
\usepackage[overload]{textcase} 

\newcommand{\cmark}{\ding{51}} % Check mark
\newcommand{\xmark}{\ding{55}} %cross mark

\def\nn{\nonumber}
\def\bea{\begin{eqnarray}}
\def\eea{\end{eqnarray}}
\def\be{\begin{equation}}
\def\ee{\end{equation}}

\def\kb{{\bm k}}

\def\e{\varepsilon}
\def\m{\mathcal}
\def\bal{\begin{aligned}}
\def\eal{\end{aligned}}
\def\bseq{\begin{subequations}}
\def\eseq{\end{subequations}}

\begin{document}

\title{Planar Hall effect in single and bilayer Rashba systems}
\author{Rahul Biswas}
% \affiliation{Department of Physics, Indian Institute of Technology, Kanpur-208016, India.}
\author{Sunit Das}
% \affiliation{Department of Physics, Indian Institute of Technology, Kanpur-208016, India.}
\author{Amit Agarwal}
\email{amitag@iitk.ac.in}
\affiliation{Department of Physics, Indian Institute of Technology, Kanpur 208016, India}

\begin{abstract}
The planar Hall effect (PHE) is an anisotropic magnetotransport response generated by coplanar electric and magnetic fields. We investigate the PHE in single- and bilayer two-dimensional electron gases (2DEGs) with Rashba spin-orbit coupling and identify two distinct mechanisms: Zeeman coupling and a band geometric channel. In the Zeeman channel, an in-plane magnetic field distorts the Rashba spin-orbit-coupled band dispersion and generates anisotropic carrier velocities, producing a finite PHE. In an asymmetric Rashba bilayer, interlayer electronic delocalization generates finite planar Berry curvature and orbital magnetic moment components, giving rise to a band geometric PHE channel. Using semiclassical Boltzmann transport theory, we calculate the chemical potential and angular dependence of the planar Hall conductivity for both mechanisms. Symmetry analysis shows that the leading response is quadratic in the magnetic field and exhibits the characteristic $\pi$-periodic angular dependence. For the parameter regime considered here, the Zeeman-induced contribution dominates, while the band geometric channel provides a distinct symmetry-allowed contribution unique to asymmetric Rashba bilayers. Our results reveal microscopic origins of anisotropic magnetotransport in spin-orbit-coupled two-dimensional materials.
\end{abstract}
\maketitle

%%%%%%%%%%%%%%%%%%%%%%%  Introduction  %%%%%%%%%%%%%%%%%%

\section{Introduction}

The planar Hall effect (PHE) refers to the appearance of a transverse voltage in the presence of coplanar electric and magnetic fields. Unlike the ordinary Hall effect, which originates from the Lorentz force, the PHE arises from anisotropic magnetoconductivity (AMC), namely the difference between conductivities measured parallel and perpendicular to the in-plane magnetic field. Such anisotropic planar transport has been observed and studied in a wide range of systems, including ferromagnets~\cite{Smit_AMR, yau1971planar, Tang_2003_Giant_PHE}, topological materials~\cite{Wang_2018_TI_PHE, taskin2017planar, Duan_TI_2020, Duan_TI_2021, TI_xiao, TI_bhardwaj, Pan_2019_TI_NPHE_experiment, Amit_NPHE_2025}, Weyl semimetals~\cite{Nitish_PHE_Weyl, Nandy_2017_PHE, PHE_weyl_liang, Amit_2022_nonlinear_magnetotransport, SA_Yang_2023_NPHE, Zyuzin_2020_phe, kamal_prb19, Azaz_prb21,PhysRevB.102.035164}, and Kramers-Weyl semimetals~\cite{Sunit_prb23, Varma_prb24}. More recently, planar Hall and planar Nernst responses originating from in-plane components of the Berry curvature (BC) and orbital magnetic moment (OMM) have been proposed in quasi-two-dimensional systems~\cite{koushik_2dphe, rahul_2dpne}.

Recently, there have been several observations of anisotropic magnetoresistance and planar Hall response in spin-orbit-coupled oxide interfaces such as $\rm LaVO_{3}$--$\rm KTaO_{3}$~\cite{wadehra_2020_phe, Tomar_prb21, Nand21, Rout_prb17}. In these materials, Rashba spin-orbit coupling naturally emerges at inversion-asymmetric oxide interfaces and strongly modifies both the electronic dispersion and spin texture~\cite{Dresselhaus2018_2DEG, mannhart2008_2DEG, Caviglia2010tunable_RSOC, Suvankar2022PHE_oxide, Amit2011SOC_spin_oscillation, sarkar2026spin, Soori2021}. Related Rashba platforms, including $\rm LaAlO_{3}$--$\rm SrTiO_{3}$ interfaces~\cite{Ariando13, yang14}, InAs- and HgTe-based quantum wells~\cite{zhang_prb04, minkov_prb20, Lei_apl19, Ashley_prb09}, and asymmetric semiconductor bilayers, also provide strong spin-orbit coupling, tunable inversion asymmetry, interlayer hybridization, or large effective $g$ factors. However, the origin of the planar Hall response in two-dimensional Rashba spin-orbit coupled systems is not fully understood~\cite{wadehra_2020_phe}. 

Motivated by this, we analyze the PHE in single- and bilayer Rashba 2DEGs and identify two distinct microscopic mechanisms. The first mechanism originates from Zeeman coupling-induced Fermi surface anisotropy. An in-plane magnetic field couples to the spin moments, distorts the electronic dispersion, and generates unequal carrier velocities parallel and perpendicular to the magnetic field. This anisotropic velocity distribution produces anisotropic magnetoconductivity and a finite planar Hall response even in the absence of BC and OMM corrections. This contribution exists in both single- and multilayer Rashba systems.

The second mechanism is of band geometric origin and can occur in quasi-two-dimensional Rashba multilayers. Interlayer electronic delocalization together with Rashba-layer asymmetry generates finite planar BC and OMM components~\cite{Jin_prb21, koushik_2dphe, rahul_2dpne, Kamal_2021_Hall_OMM}, which contribute additional planar Hall currents. We derive the corresponding planar Hall conductivities within semiclassical Boltzmann transport theory and explicitly calculate the band geometry-driven PHE in Rashba bilayers, comparing their angular and chemical potential dependences.

We perform a detailed crystalline symmetry analysis of the planar Hall response tensor and find that a large class of two-dimensional spin-orbit-coupled materials can exhibit planar Hall responses. Our analysis shows that, for the parameter regime considered here, the Zeeman-induced mechanism provides the dominant contribution to the PHE, while the BC-OMM channel gives a smaller but symmetry-allowed geometric correction unique to Rashba bilayers. These results establish a minimal framework for understanding planar Hall transport in spin-orbit-coupled oxide-interface 2DEGs and clarify the distinct roles of Rashba spin-orbit coupling, Zeeman-coupling-induced Fermi surface anisotropy, and bilayer band geometry. 

The remainder of the paper is organized as follows. In Sec.~\ref{Sec_2}, we discuss the microscopic mechanisms underlying the planar Hall effect and summarize the symmetry constraints governing the allowed planar transport responses. In Sec.~\ref{PHE in Rashba 2DEG from Zeeman}, we analyze the Zeeman-coupling-driven planar Hall effect in single-layer and bilayer Rashba 2DEGs arising from Fermi surface anisotropy. In Sec.~\ref{PHE in Rashba 2DEG from BC and OMM}, we investigate the band geometric contribution to the planar Hall effect in Rashba bilayers and identify the conditions required for finite planar Berry curvature and orbital magnetic moment. In Sec.~\ref{sec_disc}, we compare the relative magnitudes of the Zeeman and band geometric channels and discuss their experimental implications. Finally, Sec.~\ref{sec_concl} summarizes our main results and conclusions.

\section{Planar Hall Response Mechanisms} \label{Sec_2}
In general, the planar Hall response originates from anisotropic magnetoconductivity, where the conductivity acquires an angular dependence determined by the relative orientation between the electric and magnetic fields. For an electric field applied along $\hat{{x}}$ and a magnetic field directed at an angle $\theta$ with respect to the electric field in the $xy$-plane, the conductivity components parallel $(\sigma_{xx})$ and perpendicular $(\sigma_{yx})$ to the applied electric field can be written as (see Appendix \ref{detailed calculation of Zeeman coupling induced phe})~\cite{Zhong_2023_recent_PHE, PHE_AMR_Bowen}
\begin{subequations}
\label{AMC}
\begin{align}
    \sigma_{xx}&=\sigma_{\perp}+(\sigma_{\parallel}-\sigma_{\perp})\cos^{2}{\theta}~,\\
    \sigma_{yx}&=(\sigma_{\parallel}-\sigma_{\perp})\sin{\theta}\cos{\theta}~.
\end{align}
\end{subequations}
Here, $\sigma_{\parallel}$ and $\sigma_{\perp}$ denote the conductivity components parallel and perpendicular to the applied magnetic field, respectively. Equation \eqref{AMC} shows that the planar Hall conductivity (PHC), $\sigma_{yx}$, is finite when the magnetoconductivity components are anisotropic, {\it i.e.}, $(\sigma_{\parallel}-\sigma_{\perp}) \neq 0$. 
We first discuss two mechanisms for producing this AMC and, consequently, the PHE in Rashba spin-orbit-coupled systems. The first is Zeeman coupling-driven band anisotropy, while the second is of band geometric origin. We then summarize the symmetry constraints on the response tensors.

\subsection{\texorpdfstring{Zeeman coupling-driven PHE}{Zeeman-coupling-driven PHE}}

We first describe how spin-Zeeman coupling from an in-plane magnetic field induces AMC in Rashba spin-orbit-coupled 2DEGs. Rashba spin-orbit coupling (SOC) lifts the spin degeneracy of the parabolic 2DEG band, leading to two spin-split bands with opposite helicities. In these bands, the electron spin is locked to the momentum and lies in the plane, perpendicular to the momentum vector. Under an applied magnetic field $\bm{B}$, the spin moments couple through the Zeeman coupling term proportional to $\bm{\sigma}\cdot \bm{B}$, where $\bm{\sigma}$ denotes the Pauli matrices. Because Rashba SOC couples spin and orbital degrees of freedom, the Zeeman term modifies not only the spin texture but also the electronic dispersion. The band structure, therefore, becomes anisotropic, leading to unequal velocity components parallel and perpendicular to the applied magnetic field and giving rise to AMC.

To establish this, we use Boltzmann transport formalism~\cite{ashcroft2022solid}. Within the relaxation-time approximation (RTA)~\cite{Ibach2009solid}, and with an electric field along $\hat{{x}}$, the steady-state non-equilibrium distribution to linear order in the electric field is given by
\begin{equation}
    f = f_{0} - e\tau E_{x} v_{x} \frac{\partial f_{0}}{\partial \e_{\bm{k}}}~.
\end{equation}
Here, $\tau$ is a constant relaxation time, $f_{0}=(1+e^{\beta (\varepsilon_{\bm{k}}-\mu)})^{-1}$ is the equilibrium Fermi-Dirac distribution for band energy $\e_{\bm k}$, chemical potential $\mu$, and $\beta = 1/(k_B T)$. $v_{x}$ denotes the band velocity for a single band along $\hat{{x}}$. Also, $[d{\bm k}]$ is shorthand for $d^2k/4\pi^2$, and for notational simplicity, we exclude explicit band indices $n$ in these quantities. The conductivity tensor is calculated to be
\begin{equation}
\label{sigma_{ax}}
    \sigma_{ax}^{\rm Z} = -{e^{2}\tau} \sum_n\int [d {\bm k}] \, v_{a}v_{x}\frac{\partial f_{0}}{\partial \e_{\bm{k}}}~.
\end{equation}
Here, the velocity components are modified band velocity incorporating the Zeeman coupling effect. This gives the planar conductivity components parallel and perpendicular to the applied electric field of Eq.~\eqref{AMC},
where $\sigma_{\parallel}$ and $\sigma_{\perp}$ are given by
    \bea\label{sigma_para_perp}
        \sigma_{\parallel, \perp} = - {e^{2}\tau} \sum_{n}\int [d{\bm k}] \, v^{2}_{\parallel, \perp} \frac{\partial f_{0}}{\partial \e_{\bm{k}}}~, %~,\\
    \eea
Here, $v_{\parallel}$ and $v_{\perp}$ are the velocity components parallel and perpendicular to the applied magnetic field. An anisotropic band dispersion gives $v_{\parallel}\neq v_{\perp}$, and hence finite $(\sigma_{\parallel}-\sigma_{\perp})$ and PHE. For later comparison with the symmetry analysis, the Zeeman coupling-driven conductivity can also be viewed as a low-field expansion of $\sigma^{\rm Z}_{ax}$ in powers of the in-plane magnetic field (Appendix \ref{detailed calculation of Zeeman coupling induced phe}). The coefficients in this expansion have the same tensor character as other magnetic field-dependent planar transport coefficients, but their microscopic origin is the velocity anisotropy rather than BC or OMM corrections. We use these conductivity expressions to calculate the PHC in single- and bilayer Rashba 2DEGs in Sec.~\ref{PHE in Rashba 2DEG from Zeeman}.
%%%%%%%%%%%%%%%%%%%%%%%%%%%%%%%%%%
\subsection{\texorpdfstring{Band geometry-driven PHE}{band geometric PHE}}\label{band geometry_phe_mechanism}
We next discuss how the band geometric quantities, BC ($\bm{\Omega}_{\bm{k}}$) and OMM ($\bm{m}_{\bm{k}}$), give rise to PHE. In semiclassical charge transport, the wave-packet equations of motion are modified by BC and OMM~\cite{sundaram_niu_wavepacket, Chang_Niu_OMM, Amit2026quantum_geometry}. The BC and OMM in the presence of a magnetic field introduce the following modifications: (i) a correction to the band energy, $\tilde{\e}_{\bm{k}}=\e_{\bm{k}}-\bm{m}_{\bm{k}}\cdot\bm{B}$, which changes the band velocity through $\tilde{\bm{v}}_{\bm{k}}\propto \bm{\nabla}_{\bm{k}}(\bm{m}_{\bm{k}}\cdot\bm{B})$, and (ii) a BC-dependent velocity term proportional to $(\tilde{\bm{v}}_{\bm{k}}\cdot\bm{\Omega}_{\bm{k}})\bm{B}$. Together, these terms can generate AMC and PHE in three-dimensional and quasi-two-dimensional materials~\cite{koushik_2dphe}.

For a strictly 2D system, where carrier motion is confined to the plane, the in-plane components of the BC and OMM vanish (Appendix \ref{Theory of planar BC and OMM}). Therefore, BC- and OMM-driven planar Hall current is absent in a single Rashba 2DEG layer. In a quasi-2D bilayer, by contrast, interlayer electronic delocalization enables finite in-plane components of BC and OMM \cite{Jin_prb21, koushik_2dphe, rahul_2dpne}. These planar components, $\bm{\Omega}_{\bm{k}}^{\rm{Pl}}$ and $\bm{m}_{\bm{k}}^{\rm{Pl}}$, can couple to the band velocity and magnetic field and produce a PHE.

The BC- and OMM-modified charge current density is $\bm{j} = -e\sum_{n}\int [d\bm{k}] D_{\bm{k}}^{-1} \dot{\bm{r}}_{n}f_{n}$, where $-e~(e>0)$ is the electronic charge, $n$ is the band index, $\dot{\bm{r}}_{n}$ is the wave-packet velocity, $f_{n}=f_{n}(\bm r,\bm k)$ is the non-equilibrium distribution function, and $D_{\bm k}=[1+\frac{e}{\hbar}(\bm{B}\cdot\bm{\Omega_{k}})]^{-1}$ is the phase-space correction factor~\cite{xiao2010berry}. Using the RTA and keeping terms linear in the applied electric field, the band geometric contribution to the planar Hall current to first and second order in the applied magnetic field can be written as
\begin{equation}
    j^{\rm BG}_a = \tau \chi^{\rm BG}_{ab;c} E_b B_c + \tau \chi^{\rm BG}_{ab;cd} E_b B_c B_d~.
\label{planar_current_expression}
\end{equation}
Here, $\{{a,b,c,d}\}\in\{x,y\}$, $\tau$ is the relaxation time, $E_{b}$ is the electric field component, and $B_{c,d}$ are magnetic field components. The superscript BG labels the band geometric channel, for which the tensors $\chi^{\rm BG}_{ab;c}$ and $\chi^{\rm BG}_{ab;cd}$ are derived from BC and OMM corrections in Appendix \ref{calculation of BC and OMM induced response}. The allowed tensor components are dictated by symmetry, as discussed next. The BC- and OMM-induced PHC is calculated in Sec.~\ref{PHE in Rashba 2DEG from BC and OMM}.

\subsection{Symmetry constraints and angular dependence}
\label{symmetry_constraints}
Symmetry constrains the field powers and angular structures that can appear in planar transport, independent of the microscopic mechanism that generates the response. For a generic low-field planar current, we can write
\begin{equation}
    j_a=\tau {\cal K}_{ab;c}E_bB_c+\tau {\cal K}_{ab;cd}E_bB_cB_d+\mathcal{O}(B^3)~,
    \label{generic_planar_response}
\end{equation}
where $\{a,b,c,d\}\in\{x,y\}$. The coefficients ${\cal K}_{ab;c}$ and ${\cal K}_{ab;cd}$ denote generic third- and fourth-rank in-plane response tensors. For the PHC originating from band geometry in Eq.~\eqref{planar_current_expression}, these coefficients are the tensors $\chi^{\rm BG}_{ab;c}$ and $\chi^{\rm BG}_{ab;cd}$, presented in Table~\ref{table2}. For the Zeeman coupling channel, the same tensor character applies to the coefficients obtained by expanding $\sigma^{\rm Z}_{ab}$ in powers of $B$, although we keep the microscopic Zeeman calculation in the conductivity language of Sec.~\ref{PHE in Rashba 2DEG from Zeeman} and Appendix \ref{detailed calculation of Zeeman coupling induced phe}.

We now determine the allowed tensor components using Neumann's principle~\cite{newnham2005properties}. The linear-$B$ response ${\cal K}_{ab;c}$ is time-reversal ($\cal T$) odd, thus can only appear in magnetic materials. In contrast, ${\cal K}_{ab;cd}$ is $\cal T$-even tensor, which can be finite in both magnetic and nonmagnetic metals. Under a point-group operation ${\cal O}$, the response tensors transform as
\bea
{\cal K}_{a'b';c'} &=& \eta_{\cal T} \det({\cal O}) {\cal O}_{a'a} {\cal O}_{b'b} {\cal O}_{c'c} {\cal K}_{ab;c}~, \\
{\cal K}_{a'b';c'd'} &=& {\cal O}_{a'a} {\cal O}_{b'b} {\cal O}_{c'c} {\cal O}_{d'd} {\cal K}_{ab;cd}~.
\eea
Here, $\eta_{\cal T} = -1$ for magnetic point-group operations (${\cal O}={\cal RT}$) and $\eta_{\cal T} = +1$ for nonmagnetic operations (${\cal O}={\cal R}$), where ${\cal R}$ is a crystalline symmetry operation.

We work with the planar Hall geometry, where $\bm{E}=E_x\hat{{x}}$ and $\bm{B}=B(\cos\theta,\sin\theta)$. The candidate $B$-linear longitudinal coefficients are ${\cal K}_{xx;x}$ and ${\cal K}_{xx;y}$, while the candidate $B$-linear transverse coefficients are ${\cal K}_{yx;x}$ and ${\cal K}_{yx;y}$. The fourth-rank tensor is symmetric in the magnetic field indices $(c,d)$, so the candidate $B^2$ longitudinal coefficients are ${\cal K}_{xx;xx}$, ${\cal K}_{xx;yy}$, and ${\cal K}_{xx;xy}$, and the candidate $B^2$ transverse coefficients are ${\cal K}_{yx;xx}$, ${\cal K}_{yx;yy}$, and ${\cal K}_{yx;xy}$. Table \ref{2dphe_table} summarizes the symmetry-allowed tensor elements.
%%%%%%%%%%%%%
%%%%%%%%%%%%%% table 2: Point group symmetry table %%%%%%%%
\begin{table*}[t!]
\setlength{\tabcolsep}{6.5pt}
\caption{Symmetry constraints on generic $B$-linear and $B$-quadratic planar response tensors. The cross (\xmark) and tick (\cmark) indicate whether the corresponding tensor component is symmetry-forbidden or symmetry-allowed, respectively. Here, ${\m M}_a$ and ${\m C}_{na}$ denote mirror and $n$-fold rotation operations along the $a$ direction, with $a = {x, y, z}$.}
\label{2dphe_table}
\begin{tabular}{c c c c c c c c c c c c c c c c}
\hline \hline 
\noalign{\vskip 2pt}
\rm{Longitudinal } & \rm{Transverse } & ${\mathcal P}$ & ${\mathcal T}$ & $\mathcal{P}\mathcal{T}$  &  ${\cal M}_x$ & ${\cal M}_y$ & ${\cal M}_z$ & ${\cal C}_{2x}$ & ${\cal C}_{2y}$ & ${\cal C}_{2z}$ & ${\cal C}_{3z}$ & ${\cal C}_{4z}$ & ${\cal C}_{6z}$ & ${\cal S}_{4z}$  & ${\cal S}_{6z}$ \\
\noalign{\vskip 2pt}
\hline \hline 

${\cal K}_{xx;x}$ & ${\cal K}_{yx;y}$ & \cmark  & \xmark  & \xmark & \cmark &  \xmark & \xmark & \cmark & \xmark & \xmark &  \cmark  & \xmark & \xmark & \xmark & \cmark \\

${\cal K}_{xx;y}$ & ${\cal K}_{yx;x}$ & \cmark  & \xmark & \xmark  & \xmark &  \cmark & \xmark & \xmark & \cmark  & \xmark & \cmark  & \xmark & \xmark & \xmark & \cmark \\

\noalign{\vskip 2pt}
\hline
\noalign{\vskip 1pt}

${\cal K}_{xx;xx}$, ${\cal K}_{xx;yy}$ & ${\cal K}_{yx;xy}$ &  \cmark & \cmark & \cmark & \cmark & \cmark & \cmark &  \cmark &  \cmark &  \cmark &  \cmark &  \cmark &  \cmark &  \cmark &  \cmark \\ 

${\cal K}_{xx;xy}$ & ${\cal K}_{yx;xx}$, ${\cal K}_{yx;yy}$ &  \cmark & \cmark & \cmark & \xmark & \xmark & \cmark & \xmark & \xmark & \cmark & \cmark & \cmark & \cmark & \cmark & \cmark \\

\noalign{\vskip 2pt}
\hline \hline
\end{tabular}
\end{table*}
%%%%%%%%%%%%%%%%%%%%%%%%%

Using Eq.~\eqref{generic_planar_response} and the planar Hall geometry, the angular dependence of the conductivities is obtained to be
\bea \label{sigma_long}
\sigma_{xx} &=& \tau B({\cal K}_{xx;x} \cos\theta + {\cal K}_{xx;y}\sin\theta) + \tau B^2 ({\cal K}_{xx;xx} \cos^2{\theta} \nn \\
&& +~{\cal K}_{xx;yy} \sin^2{\theta} + {\cal K}_{xx;xy}\sin{\theta}\cos{\theta})~,
\eea
\bea \label{sigma_trans}
\sigma_{yx} &=& \tau B ({\cal K}_{yx;x} \cos{\theta}+{\cal K}_{yx;y} \sin{\theta}) + \tau B^2 ({\cal K}_{yx;xx} \cos^2{\theta} \nn \\
&& +~{\cal K}_{yx;yy} \sin^2{\theta} + {\cal K}_{yx;xy}\sin{\theta}\cos{\theta})~. \label{sigma_hall}
\eea
For a response channel whose symmetry contains both $\mathcal{M}_{x}$ and $\mathcal{M}_{y}$, the third-rank terms are forbidden, and $\mathcal{M}_{x}$ and $\mathcal{M}_{y}$ also set ${\cal K}_{xx;xy}$, ${\cal K}_{yx;xx}$, and ${\cal K}_{yx;yy}$ to zero (see Table~\ref{2dphe_table}). In the presence of these symmetries, the leading symmetry-allowed angular dependence then reduces to
\begin{align}
\sigma_{xx}^{(2)} &= \tau B^{2}
\left({\cal K}_{xx;xx}\cos^{2}{\theta}+{\cal K}_{xx;yy}\sin^{2}{\theta}\right)~,\\
\sigma_{yx}^{(2)} &= \tau B^{2}{\cal K}_{yx;xy}\sin{\theta}\cos{\theta}~.
\end{align}
Symmetry, therefore, characterizes the possible tensor structures and angular forms, while the microscopic calculation determines whether an allowed coefficient is nonzero and how large it is. In the following sections, this framework is first applied to the Zeeman-driven Rashba velocity-anisotropy channel and then to the band geometric BC-OMM channel in a Rashba bilayer.

%%%%%%%%%%%%%%%
\begin{figure}
    \centering  \includegraphics[width=\linewidth]{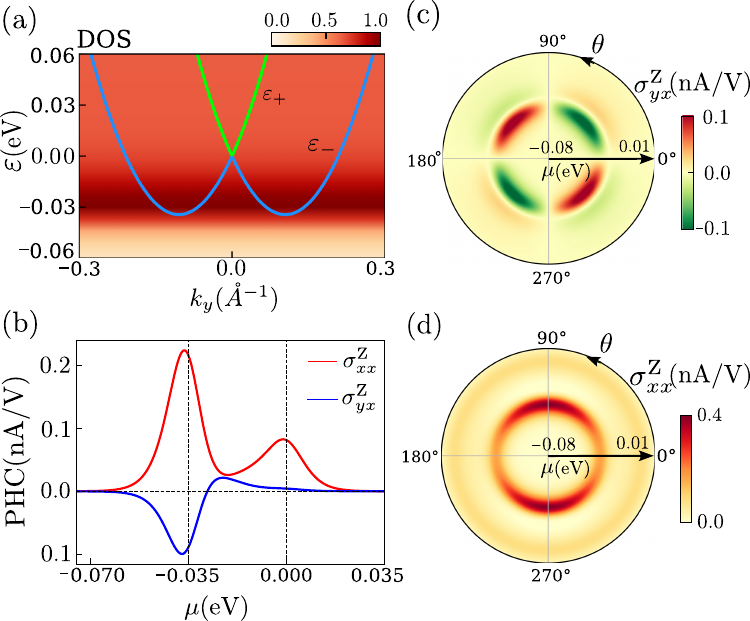}
    \caption{Zeeman coupling induced PHE in a single-layer Rashba 2DEG. (a) Electronic band dispersion, with the density of states (DOS) indicated by the background color. (b) Variation of the longitudinal and transverse planar conductivities, $\sigma^{\rm Z}_{xx}$ and $\sigma^{\rm Z}_{yx}$ as functions of $\mu$ for fixed angle $\theta = 30^{\circ}$. For $\sigma^{\rm Z}_{xx}$, only the magnetic field-dependent contribution is shown; the Drude background is not included. (c,d) Angular dependence of $\sigma^{\rm Z}_{yx}$ and $\sigma^{\rm Z}_{xx}$, as the chemical potential $\mu$ varies. We use effective mass $m=1.2~m_{e}$, Rashba parameter $\alpha=0.66$ eV$\text{\AA}$, magnetic field $B=0.3$ T, temperature $T=50$ K, and relaxation time $\tau=1$ ps.}
    \label{dispersion_}
\end{figure}
%%%%%%%%%%%%%%%%%%%

%%%%%%%%%%%%%%%%
\section{Zeeman-driven PHE in Rashba 2DEGs}
\label{PHE in Rashba 2DEG from Zeeman}
We first calculate the PHE generated by Zeeman-induced Fermi surface anisotropy in single- and bilayer Rashba 2DEGs.
Our motivation comes from Rashba 2DEGs realized in oxide heterostructures and interfaces, where spin-orbit-coupled electronic states can appear as a single interfacial 2DEG or as coupled Rashba-like layers~\cite{Dresselhaus2018_2DEG, mannhart2008_2DEG, Caviglia2010tunable_RSOC, Suvankar2022PHE_oxide}. We therefore use simplified monolayer and bilayer Rashba-2DEG descriptions.

For the single-layer Hamiltonian and the bilayer Hamiltonian used below, the parabolic term, Rashba SOC term, and Zeeman term remain invariant under the in-plane mirror operations $\mathcal{M}_{x}$ and $\mathcal{M}_{y}$. These Rashba symmetries are the model-specific input for applying the tensor constraints of Sec.~\ref{symmetry_constraints}; in particular, the simultaneous presence of $\mathcal{M}_{x}$ and $\mathcal{M}_{y}$ forbids the $B$-linear transverse response tensor (see Table~\ref{2dphe_table}).
\subsection{Single-layer Rashba 2DEG}
We begin with the Hamiltonian of a single 2DEG with Rashba SOC and Zeeman coupling~\cite{Rashba_1984, soumya2016_SOCinterface, Eremeev2012_2DEG_RSOC},
\begin{equation}
\label{Hamiltonian}
\mathcal{H}_{s} = \frac{\hbar^{2}k^{2}}{2m} \sigma_0 + \alpha\bm{\sigma}\cdot(\bm k \times {\hat{z}})+g' \bm{\sigma}\cdot \bm{B}~, 
\end{equation}
where $\bm{k}$ is the Bloch wave vector, $m$ is the effective electron mass, $\alpha$ is the Rashba SOC parameter, and $\bm{\sigma}$ represents the Pauli matrices acting in spin space. The second term is the Rashba SOC, and the third term is the Zeeman interaction, with $g'=g\mu_{B}/2$, effective $g$-factor $g$, and electron Bohr magneton $\mu_{B}=e\hbar/(2m)$. The energy eigenvalues are
\begin{equation}
\label{eigenvalue}
    \e_{\pm} = \frac{\hbar^{2}k^{2}}{2m} \pm \sqrt{2\alpha g'(k_{y}B_{x}-k_{x}B_{y}) + \alpha^{2}k^{2}+ g'^{2}B^{2}}~.
\end{equation}
The anisotropy of the band velocities becomes explicit when the dispersion is expressed in terms of components parallel and perpendicular to the applied magnetic field [Eq.~\eqref{E_rotated}]. Figure~\ref{dispersion_}(a) shows the band dispersion of Hamiltonian~\eqref{Hamiltonian} together with the electronic density of states. The upper band (green) contains only positive energies, whereas the lower band (blue) has both positive and negative energies. The density of states (DOS) has a jump near the band edge around $\mu=-0.028$ eV.

Equation \eqref{AMC} governs the angular dependence of the Zeeman-driven planar Hall response. The transverse conductivity $\sigma^{\rm Z}_{yx}$ contains only $B$-dependent contributions, whereas the longitudinal conductivity $\sigma^{\rm{Z}}_{xx}$ contains both $B$-dependent terms and the $B$-independent Drude contribution. This follows from the expansion in powers of $\bm{B}$ [Eq. \eqref{Zeeman_sig1}]. In that expansion, the $B^0$ and $B^1$ terms in $(\sigma_{\parallel}-\sigma_{\perp})$ vanish after momentum-space integration, so the leading Zeeman-coupling driven PHC for the Rashba model considered here is quadratic in $B$ and has the form $\sigma^{\rm Z}_{yx}=\tau B^2 \mathcal{K}_{yx;xy} \sin\theta\cos\theta$. In our numerical results, we exclude the conventional Drude contribution from the Zeeman-driven longitudinal conductivity $\sigma^{\rm Z}_{xx}$.

Figure \ref{dispersion_}(b) shows the magnetic field-dependent longitudinal and transverse PHE conductivities, $\sigma^{\rm Z}_{xx}$ and $\sigma^{\rm Z}_{yx}$, as functions of chemical potential for a single-layer Rashba 2DEG at fixed angle $\theta=30^{\circ}$ between the electric and magnetic fields. Near the band edges, the derivative of the Fermi function, $\partial f_{0}/\partial \e_{\bm{k}}$, becomes sharply peaked and enhances both conductivities. At $\mu=-0.028$ eV, $\sigma^{\rm Z}_{yx}$ changes sign and then gradually approaches zero with increasing $\mu$. Figures \ref{dispersion_}(c,d) show the angular variation of $\sigma^{\rm Z}_{yx}$ and $\sigma^{\rm Z}_{xx}$ with varying chemical potential. The chemical potential is $\mu=-0.08$ eV at the origin of the polar plot and increases radially to $0.01$ eV; the color represents the conductivity in nA/V. The transverse response follows a characteristic $\sin{2\theta}$ angular dependence, whereas the longitudinal response varies as $\cos^{2}{\theta}$. Thus, Zeeman coupling drives anisotropic magnetoconductivity and a planar Hall response with conventional angular characteristics in single-layer Rashba SOC systems.

\subsection{Bilayer Rashba 2DEG}

\begin{figure}
    \centering  \includegraphics[width=\linewidth]{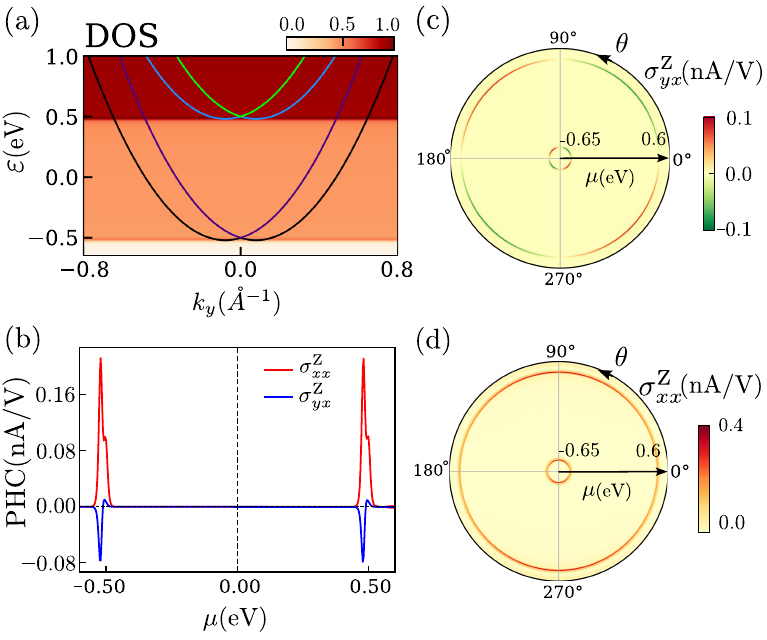}
    \caption{Zeeman-driven PHE in a bilayer Rashba 2DEG. (a) Electronic band dispersion, with the density of states (DOS) indicated by the background color. (b) Variation of the longitudinal and transverse conductivities, $\sigma^{\rm Z}_{xx}$ and $\sigma^{\rm Z}_{yx}$ as functions of $\mu$ for fixed angle $\theta = 30^{\circ}$. The two response peaks occur near the band edges of the two Rashba layers. For $\sigma^{\rm Z}_{xx}$, only the magnetic field-dependent contribution is shown. (c,d) Angular dependence of $\sigma^{\rm Z}_{yx}$ and $\sigma^{\rm Z}_{xx}$, as the chemical potential $\mu$ varies. We use equal effective masses, $m=1.2~m_{e}$, Rashba parameters $\alpha_{A}=2\alpha_{B}=0.66$ eV$\text{\AA}$, interlayer hopping $t=0.5$ eV, magnetic field $B=0.3$ T, temperature $T=50$ K, and relaxation time $\tau=1$ ps.}
    \label{dispersion_bilayers}
\end{figure}

We now consider a bilayer system consisting of two 2DEG layers separated by a thin barrier. The Hamiltonian of a bilayer 2DEG with Rashba SOC and Zeeman coupling is~\cite{Sergio2023_spin_Edelstein, tanmoy2013TI_Rashba_bilayers, bianchi2010_2DEG_TISS, michiardi2022optical_manipulation_RSOC}
\begin{equation}\label{Hamiltonian_bilayers}
    \mathcal{H}_{b}= \begin{pmatrix}
        \mathcal{H}_{A} & T\\
        T & \mathcal{H}_{B}
    \end{pmatrix}~,  ~~~~~   T=\begin{pmatrix}
        t & 0 \\
        0 & t
    \end{pmatrix}~,
\end{equation}
with $\mathcal{H}_{A},~\mathcal{H}_{B}$ expressed as
\begin{equation}
\mathcal{H}_{l}=\frac{\hbar^{2} \bm k^{2}}{2m_{l}} \sigma_0 + \alpha_{l} \bm{\sigma} \cdot (\bm k \times \hat{z})+g'\bm{\sigma}\cdot \bm{B}~, ~~~l=A,B.
\end{equation}
The spin-independent interaction between the two layers is described by the hopping matrix $T$, where $t$ denotes the interlayer hopping strength. The electronic band dispersion of the bilayer Hamiltonian \eqref{Hamiltonian_bilayers}, shown in Fig. \ref{dispersion_bilayers}(a), has four Rashba spin-split bands, two from the lower 2DEG layer and two from the upper 2DEG layer. The DOS peaks around the band edges are indicated by the background color.

Figure \ref{dispersion_bilayers}(b) shows the longitudinal and transverse PHE conductivities of a Zeeman-coupled bilayer Rashba 2DEG as functions of chemical potential at fixed angle $\theta=30^{\circ}$. The conductivities show two peaks near the band edges of the two Rashba 2DEG layers, around $\mu=-0.5$ eV and $\mu=0.5$ eV. Figures \ref{dispersion_bilayers}(c,d) show the angular dependence of $\sigma^{\rm Z}_{xx}$ and $\sigma^{\rm Z}_{yx}$ as $\mu$ increases radially. At the center of the polar plot, $\mu=-0.65$ eV, and it increases radially to $0.6$ eV. 
%
%%%%%%%%%%%%%%%%%%%%%%
%%%%%%%%%%%%%%%%%%%%%%% 

\section{Band geometric PHE in Rashba bilayers}
\label{PHE in Rashba 2DEG from BC and OMM}
We next estimate the BC- and OMM-driven PHE in a Rashba bilayer. As discussed in Sec.~\ref{band geometry_phe_mechanism}, this mechanism is absent in a strictly single-layer Rashba 2DEG because the planar BC and OMM components vanish when the electronic motion is confined to the plane. A coupled bilayer provides the minimal extension in which the wave functions have layer-dependent out-of-plane structure. Interlayer electronic delocalization can then generate finite planar BC and OMM components, allowing a distinct band geometric PHE channel.

%%%%%%%%%%%%%%%
\begin{figure*}[t!]
    \centering
    \includegraphics[width=0.95\linewidth]{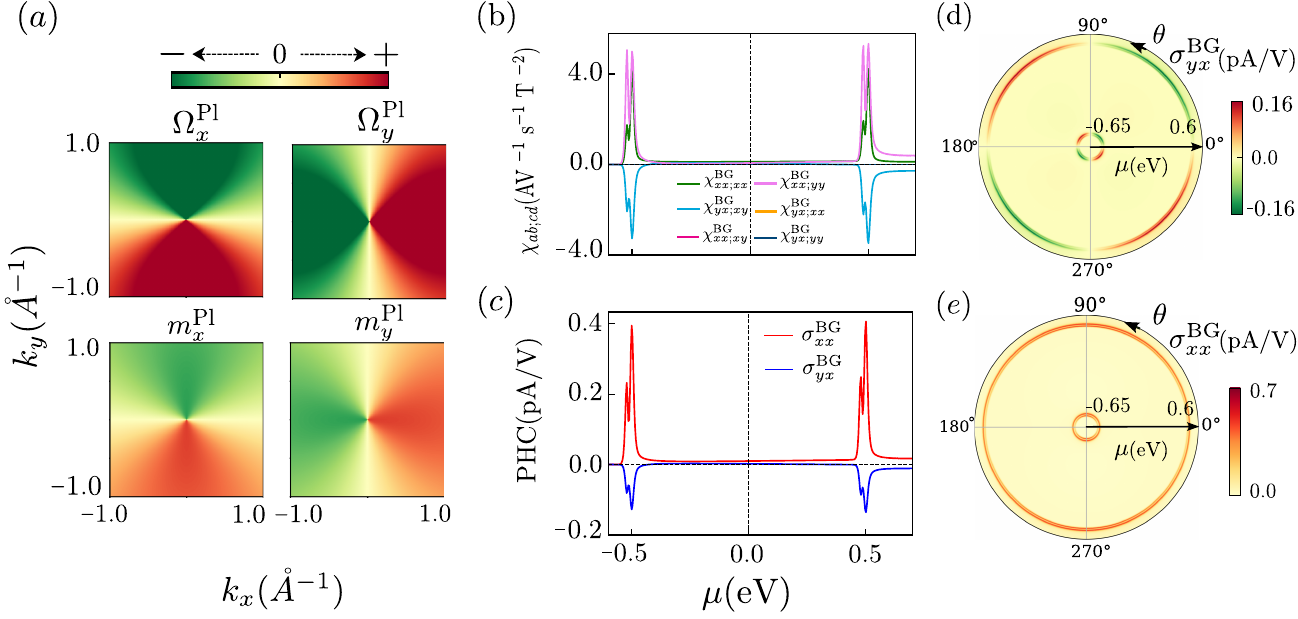}
    \caption{Band geometric PHE in a Rashba bilayer.  (a) Momentum-space distribution of the planar BC ($\Omega^{\rm Pl}$) and OMM ($m^{\rm Pl}$) components for the lowest band. (b) BC and OMM driven $B^{2}$-dependent planar responses as functions of chemical potential. (c) Longitudinal ($\sigma^{\rm BG}_{xx}$) and transverse ($\sigma^{\rm BG}_{yx}$) planar Hall conductivities for fixed angle $\theta=30^{\circ}$. The largest features occur near band edges, where the planar BC and OMM are enhanced. (d,e) Angular dependence of the planar Hall conductivities as $\mu$ varies. The transverse conductivity follows $\sigma^{\rm BG}_{yx}\propto \chi^{\rm BG}_{yx;xy}\sin{\theta}\cos{\theta}$ because $\chi^{\rm BG}_{yx;xy}$ is the dominant transverse tensor, whereas the longitudinal conductivity is dominated by $\chi^{\rm BG}_{xx;xx}$ and $\chi^{\rm BG}_{xx;yy}$ and varies as $\sigma^{\rm BG}_{xx}\propto \chi^{\rm BG}_{xx;xx}\cos^{2}{\theta}+\chi^{\rm BG}_{xx;yy}\sin^{2}{\theta}$. We use effective masses $m_{A}=m_{B}=1.2~m_{e}$, Rashba parameters $\alpha_{A}=2\alpha_{B}=0.66$ eV$\text{\AA}$, interlayer hopping $t=0.5$ eV, magnetic field $B=0.3$ T, relaxation time $\tau = 1$ ps, and temperature $T=50$ K.}
\label{bc_conductivity_plot}
\end{figure*}
%%%%%%%%%%%%%%%

The planar BC and OMM are controlled by both interlayer hybridization and the symmetry relation between the two Rashba layers. For equal effective masses and identical Rashba couplings, $\alpha_A=\alpha_B$, the bilayer Hamiltonian is invariant under layer exchange, $
[{\cal H}_b, {\cal P}_L]=0$,
where ${\cal P}_L$ denotes the layer-exchange operator. The eigenstates can therefore be chosen to have definite layer parity. In this basis, the in-plane velocity operator is even under layer exchange, $[\partial_{\bm k} {\cal H},{\cal P}_L]=0$, whereas the out-of-plane position operator entering Eqs.~\eqref{Omega_pl} and~\eqref{m_pl} is odd,
$\{{\cal Z},{\cal P}_L\}=0$. Consequently, the matrix element $\langle u^n_{\bm k}|\partial_{\bm k} {\cal H}|u_{\bm k}^m\rangle$ is nonzero only between states with the same layer parity, while $\langle u^m_{\bm k}|{\cal Z}| u^n_{\bm k}\rangle$ is nonzero only between states with opposite layer parity. Since the planar BC and OMM are constructed from products of the form $\langle u^n_{\bm k}|\partial_{\bm k} {\cal H}|u^m_{\bm k} \rangle \langle u^m_{\bm k}|{\cal Z}|u^n_{\bm k}\rangle$, the two selection rules cannot be satisfied simultaneously, forcing $\bm{\Omega}^{\rm Pl}_{n\bm k}=\bm{m}^{\rm Pl}_{n\bm k}=0$. This is consistent with the fact that $\bm{\Omega}^{\rm Pl}_{n\bm k}$ and $\bm{m}^{\rm Pl}_{n\bm k}$ require interlayer coherence~\cite{koushik_2dphe, rahul_2dpne, Zheng2025}.

For equal effective masses and opposite Rashba couplings, $\alpha_A=-\alpha_B$, the unbiased bilayer possesses inversion symmetry. Since the band geometric contribution is evaluated in the absence of the Zeeman coupling, the Hamiltonian also preserves time-reversal symmetry. The simultaneous presence of inversion and time-reversal symmetries forces the planar Berry curvature and orbital magnetic moment to vanish identically~\cite{koushik_2dphe}. 

A finite planar Berry curvature or orbital magnetic moment, therefore, requires breaking inversion symmetry. This can be achieved, for example, through unequal Rashba couplings, unequal effective masses, or an interlayer potential bias. We focus on the minimal asymmetric case with equal effective masses and unequal Rashba couplings, $\alpha_A=2\alpha_B$, which generates finite planar Berry curvature and orbital magnetic moment.

Figure~\ref{bc_conductivity_plot}(a) shows the resulting momentum-space distribution of the planar BC and OMM for the lowest band. The $B^2$ band geometric response tensors, $\chi^{\rm BG}_{xx;xx}$, $\chi^{\rm BG}_{xx;yy}$, and $\chi^{\rm BG}_{yx;xy}$, are plotted as functions of chemical potential in Fig.~\ref{bc_conductivity_plot}(b). These tensors are enhanced near the bilayer band edges, where the interband denominators entering the planar BC and OMM become small. Figure~\ref{bc_conductivity_plot}(c) shows the corresponding longitudinal and transverse planar Hall conductivities at fixed angle $\theta=30^{\circ}$. Their peaks follow the same band-edge enhancement. Figures~\ref{bc_conductivity_plot}(d,e) show the angular dependence. The transverse response follows the allowed $B^2$ tensor structure and varies as $\sin\theta\cos\theta$, while the longitudinal response is dominated by the corresponding $\cos^2\theta$ and $\sin^2\theta$ tensor components. Thus, the band geometric PHE in this model is a bilayer-specific response controlled by interlayer delocalization of electrons, Rashba-layer asymmetry, and the proximity of the chemical potential to the band edge-enhanced planar BC and OMM.

%%%%%%%%%%%%%%%%%%
\section{Discussion}\label{sec_disc}

For the symmetry setting considered here, the leading nonvanishing transverse response has the same $B^2\sin\theta\cos\theta$ angular form for both mechanisms. This common dependence, however, does not imply comparable magnitudes. In the weak-field regime, the transverse conductivities can be written schematically as
$\sigma^{\rm Z}_{yx}=\tau B^2\Lambda_{\rm Z}(\mu)\sin\theta\cos\theta$
and
$\sigma^{\rm BG}_{yx}=\tau B^2\chi^{\rm BG}_{yx;xy}(\mu)\sin\theta\cos\theta$.
Here, $\Lambda_{\rm Z}(\mu)$ is the coefficient of the $B^{2}$ part of $(\sigma_{\parallel}-\sigma_{\perp})/\tau$ generated by the Zeeman-induced velocity anisotropy, and $\chi_{yx;xy}^{\rm BG}(\mu)$ is the response tensor arising from the band geometry presented in Table~\ref{table2}. The factors $\tau$, $B^2$, and $\sin\theta\cos\theta$ are fixed by symmetry and by the weak-field expansion. The relative strength of the two channels is therefore controlled by the microscopic coefficients: the Zeeman-induced Fermi surface anisotropy coefficient $\Lambda_{\rm Z}$ and the band geometric tensor $\chi^{\rm BG}_{yx;xy}$, as discussed in Appendix~\ref{relative_scale_estimate}.

To estimate the corresponding transverse voltage, we use
\begin{equation}
     V_{\perp}
     \approx
     \left\vert
     \frac{j^{\rm PHE}_{\perp}W}{\sigma^{\rm D}_{xx}}
     \right\vert
     =
     \left\vert
     \frac{\sigma_{yx}EW}{\sigma^{\rm D}_{xx}}
     \right\vert ,
\end{equation}
valid in the small Hall-angle limit, $\sigma_{yx}\ll \sigma^{\rm D}_{xx}$. Here, $W$ is the sample width, $E$ is the applied longitudinal electric field, and $\sigma^{\rm D}_{xx}$ is the longitudinal Drude conductivity. For an order-of-magnitude estimate, we take $W=10~\mu{\rm m}$, $E=1~{\rm V}/\mu{\rm m}$, and $T=50$ K. For the planar Hall conductivity, we use $\mu=0.48$ eV, $B=1$ T, $\tau=1$ ps, and $\theta=30^{\circ}$. This chemical potential lies near a band-edge-enhanced region of the bilayer response in Figs.~\ref{dispersion_bilayers} and~\ref{bc_conductivity_plot}. We obtain $V^{\rm Z}_{\perp}\approx 22.48$ nV for the Zeeman-driven response and $V^{\rm BG}_{\perp}\approx 0.03$ nV for the BC-OMM response.

For these parameters, the Zeeman channel is dominant because the field-induced anisotropic distortion of the Rashba Fermi surface produces a much larger coefficient, $|\Lambda_{\rm Z}|\gg |\chi^{\rm BG}_{yx;xy}|$. This numerical hierarchy is not universal; it identifies the dominant mechanism only for the present minimal Rashba-bilayer model. The BC-OMM channel is smaller here because it relies on several additional ingredients: interlayer delocalization, Rashba-layer asymmetry, finite planar BC and OMM, and the corresponding OMM- and BC-induced corrections to the semiclassical current in Eq.~\eqref{current_final}.

More importantly, the two channels can be separated experimentally in a controlled comparison because they rely on different microscopic origins. The Zeeman contribution requires Rashba spin-momentum locking and an in-plane magnetic field. It is finite even in a single Rashba layer and remains finite in symmetric bilayers. In contrast, the BC-OMM contribution is intrinsically a multilayer band geometric effect. It requires interlayer coupling, finite planar BC and OMM, and broken layer symmetry. It vanishes in a strictly single layer and is strongly suppressed in inversion-symmetric bilayers. Comparing the planar Hall response of a single layer, a symmetric bilayer, and an asymmetric bilayer can help separate the band geometric contribution from the Zeeman background, provided disorder, carrier density, and relaxation-time changes are controlled. In addition, an interlayer bias or tunable Rashba-layer asymmetry should strongly modify the BC-OMM contribution through changes in the planar BC and OMM, while leaving the Zeeman mechanism comparatively less affected. 
%\textcolor{red}{A systematic dependence of the planar Hall signal on layer asymmetry would therefore provide a direct signature of the band-geometric planar Hall channel.}

The ingredients underlying the two mechanisms discussed here are available in several existing material platforms. Rashba oxide interfaces, such as LaVO$_3$--KTaO$_3$, naturally host interfacial spin-orbit coupling and have already exhibited anisotropic magnetotransport and planar Hall responses. Semiconductor quantum wells, including InAs- and HgTe-based structures, provide another attractive platform because they combine strong spin-orbit coupling with large effective $g$ factors. While the present theory is not intended as a material-specific description of these systems, the coexistence of sizable spin-orbit and Zeeman energy scales makes them natural candidates for exploring the interplay between Zeeman-induced and band geometric contributions to planar Hall transport.

% \textcolor{red}{The ingredients required for observing these channels are available in several existing material platforms. Rashba oxide interfaces such as LaVO$_3$/KTaO$_3$ and LaAlO$_3$/SrTiO$_3$ naturally provide interfacial spin-orbit coupling and have already shown anisotropic magnetotransport and planar Hall responses. Semiconductor quantum wells, including InAs- and HgTe-based structures, provide another promising route because they combine tunable Rashba coupling with large effective $g$ factors. These platforms therefore provide realistic settings for separating the Zeeman-driven and band geometric planar Hall mechanisms.}

\section{Conclusion}\label{sec_concl}

In summary, we have investigated the planar Hall effect in Rashba spin-orbit-coupled two-dimensional electron gases and identified two distinct microscopic mechanisms. The first originates from Zeeman coupling-induced Fermi surface anisotropy: an in-plane magnetic field distorts the band dispersion, generates anisotropic carrier velocities, and produces a finite planar Hall response. This mechanism operates in both single- and bilayer Rashba systems. The second mechanism is specific to asymmetric Rashba bilayers, where interlayer electronic delocalization and broken inversion symmetry generate finite planar Berry curvature and orbital magnetic moment components, giving rise to an additional band geometric contribution to the planar Hall conductivity.

Our symmetry analysis shows that both mechanisms produce a leading planar Hall response that is quadratic in the in-plane magnetic field and has the characteristic $\sin\theta\cos\theta$ angular dependence. Thus, the same angular and magnetic-field dependence can arise from two different microscopic origins. This suggests that controlled comparison across single-layer, symmetric-bilayer, and asymmetric-bilayer Rashba systems can be used to separate the Zeeman-coupling induced and band geometric contributions. 
These results provide a minimal framework for understanding and separating planar Hall mechanisms in Rashba spin-orbit-coupled systems. Relevant platforms include oxide interfaces~\cite{wadehra_2020_phe, Tomar_prb21}, InAs- and HgTe-based quantum wells with strong spin-orbit coupling and large effective $g$-factors~\cite{zhang_prb04, minkov_prb20, Lei_apl19, Ashley_prb09}, as well as asymmetric semiconductor bilayers and gate-tunable Rashba multilayers.

\section*{Acknowledgements}
R. B. acknowledges the Indian Institute of Technology Kanpur for the Ph.D. fellowship.
%%%%%%%%%%%%%%%%%%%%%%%%

% \onecolumngrid
\appendix
\section{Detailed calculation of Zeeman coupling-induced PHE}
\label{detailed calculation of Zeeman coupling induced phe}

\begin{figure}
    \centering
    \includegraphics[width=\linewidth]{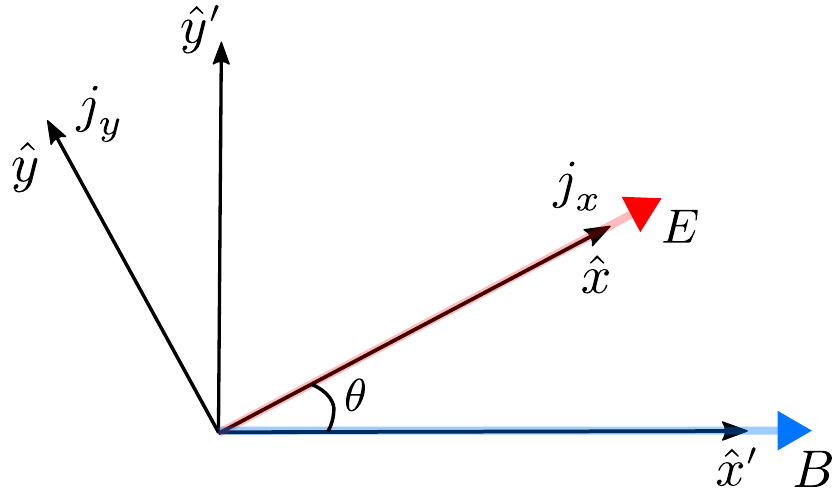}
    \caption{Rotated coordinate system used to derive the AMC form of the PHE. The electric field $\bm{E}$ is along $\hat{x}$, the magnetic field $\bm{B}$ is along $\hat{x}'$, and the current components $j_{x}$ and $j_{y}$ are parallel and perpendicular to $\bm{E}$, respectively.}
    \label{coordinate}
\end{figure}

In this Appendix, we show how magnetic field-induced Zeeman coupling produces anisotropic band velocities and generates the PHE. We first recall how the PHE follows from AMC. We consider a coordinate system where $\bm{B}$ is applied along $\hat{x}'$ and $\bm{E}$ is applied at an angle $\theta$ with respect to $\hat{x}'$, as shown in Fig.~\ref{coordinate}. The current density, $j'_{a}=\sigma'_{ab}E'_{b}$, then has the form
\begin{equation}
   \begin{pmatrix}
       j_{x}'\\
       j_{y}'
   \end{pmatrix}
   =
   \begin{pmatrix}
\sigma_{\parallel} & 0 \\
0 & \sigma_{\perp}
\end{pmatrix}
\begin{pmatrix}
    E'_{x}\\
    E'_{y}
\end{pmatrix}~,
\end{equation}
where $\sigma_{\parallel}$ and $\sigma_{\perp}$ are the conductivity components parallel and perpendicular to $\bm{B}$, respectively. We rotate the coordinate system so that the new $\hat{x}$ axis coincides with $\bm{E}$ and the new $\hat{y}$ axis is perpendicular to $\bm{E}$. The conductivity matrix in the new coordinate system is $\sigma=R\sigma'R^{T}$, where $R$ is the two-dimensional rotation matrix. This gives
\begin{widetext}
\begin{eqnarray}
    \begin{pmatrix}
        \sigma_{xx} & \sigma_{xy} \\
        \sigma_{yx} & \sigma_{yy}
    \end{pmatrix}
    &=& 
    \begin{pmatrix}
\cos \theta & -\sin \theta \\
\sin \theta & \cos \theta
\end{pmatrix}
\begin{pmatrix}
\sigma_{\parallel} & 0 \\
0 & \sigma_{\perp}
\end{pmatrix}
\begin{pmatrix}
\cos \theta & \sin \theta \\
-\sin \theta & \cos \theta
\end{pmatrix}~,\\
\Rightarrow
\begin{pmatrix}
        \sigma_{xx} & \sigma_{xy} \\
        \sigma_{yx} & \sigma_{yy}
    \end{pmatrix}
    &=&
    \begin{pmatrix}
\sigma_{\perp}+(\sigma_{\parallel}-\sigma_{\perp})\cos^{2}{\theta} &
(\sigma_{\parallel}-\sigma_{\perp})\sin{\theta}\cos{\theta} \\
(\sigma_{\parallel}-\sigma_{\perp})\sin{\theta}\cos{\theta} &
\sigma_{\perp}+(\sigma_{\parallel}-\sigma_{\perp})\sin^{2}{\theta}
\end{pmatrix}~.
\end{eqnarray}
\end{widetext}
With $\bm{E}=E\hat{x}$, the current densities parallel and perpendicular to the applied electric field are $j_{x}=\sigma_{xx}E_x$ and $j_{y}=\sigma_{yx}E_x$, respectively. Thus,
\begin{eqnarray}
\label{sig_yx_appendix}
    \sigma_{xx} &=& \sigma_{\perp}+(\sigma_{\parallel}-\sigma_{\perp})\cos^{2}{\theta} ~,\\
    \sigma_{yx} &=& (\sigma_{\parallel}-\sigma_{\perp})\sin{\theta}\cos{\theta}~,
\end{eqnarray}
where $\sigma_{\parallel}$ and $\sigma_{\perp}$ are given in Eq.~\eqref{sigma_para_perp} of the main text. We define $k_{\parallel}$ ($k_{\perp}$) and $v_{\parallel}$ ($v_{\perp}$) as the components parallel (perpendicular) to the applied magnetic field, so that $(k_{\parallel},k_{\perp})^{T}=R(k_{x},k_{y})^{T}$ and $(v_{\parallel},v_{\perp})^{T}=R(v_{x},v_{y})^{T}$, where $T$ denotes the transpose.

Next, we calculate these conductivities for a single layer of Zeeman-coupled Rashba 2DEG. In terms of $k_{\parallel}$ and $k_{\perp}$, the energy eigenvalues in Eq.~\eqref{eigenvalue} become
\begin{equation}
\label{E_rotated}
    \varepsilon_{\lambda\bm{k}}
    =
    \frac{\hbar^{2}k^{2}}{2m}
    +\lambda \Delta_{\bm k} ~.
    %\sqrt{2\alpha k_{\perp}g' B+\alpha^{2}k^{2}+g'^{2}B^{2}}~,
\end{equation}
Here, $\lambda=\pm$ labels the two Rashba branches,  $k^{2}=k^{2}_{\parallel}+k^{2}_{\perp}=k^{2}_{x}+k^{2}_{y}$, and we have defined $\Delta_{\bm k}=\sqrt{2\alpha k_{\perp}g'B+\alpha^{2}k^{2}+g'^{2}B^{2}}$. The corresponding velocities are
\begin{align}
    v_{\lambda,\parallel}
    &=
    \frac{1}{\hbar}\frac{\partial \varepsilon_{\lambda\bm{k}}}{\partial k_{\parallel}}
    =
    \frac{\hbar k_{\parallel}}{m}
    +\lambda
    \frac{\alpha^{2}k_{\parallel}}
    {\hbar\Delta_{\bm k}}~, \nonumber\\
    v_{\lambda,\perp}
    &=
    \frac{1}{\hbar}\frac{\partial \varepsilon_{\lambda\bm{k}}}{\partial k_{\perp}}
    =
    \frac{\hbar k_{\perp}}{m}
    +\lambda\frac{\alpha^{2}k_{\perp}+\alpha g'B}
    {\hbar\Delta_{\bm k}}~.
\end{align}
The Zeeman term makes the band velocities anisotropic. This generates a finite $(\sigma_{\parallel}-\sigma_{\perp})$ and, consequently, a planar Hall response in the Rashba 2DEG. 

The Zeeman coupling-induced planar conductivity, denoted by $\sigma^{\rm Z}_{yx}$, can be obtained from $\sigma_\parallel-\sigma_\perp$. It can be expanded in powers of $B$ in the low-field regime as
\begin{widetext}
\begin{eqnarray}
\label{Zeeman_sig1}
\sigma_{\parallel}-\sigma_{\perp}
&=&
- \frac{e^{2}\tau}{4 \pi^{2}}
\sum_{\lambda=\pm}
\int d^{2}\bm{k}
\biggl[
\frac{(4k^{2}+m^{2}\alpha^{2}+4m\alpha k)(k^{2}_{\parallel}-k^{2}_{\perp})}{m^{2}k^{2}}
-\frac{4g' k^{2}_{\parallel}k_{\perp}(2k^{2}+\alpha m k)}{mk^{5}}B
\nn\\
&&
-\frac{2g'^{2}k^{2}_{\parallel}
\bigl(k^{4}_{\parallel}-4k^{2}_{\parallel}k^{2}_{\perp}-5k^{4}_{\perp}
+m\alpha k(k^{2}_{\parallel}-3k^{2}_{\perp})\bigr)}
{m\alpha k^{7}}B^{2}
+\mathcal{O}(B^{3})
\biggr]
\frac{\partial f_{0}}{\partial \varepsilon}~, \\
\sigma_{\perp}
&=&
- \frac{e^{2}\tau}{4 \pi^{2}}
\sum_{\lambda=\pm}
\int d^{2}\bm{k}
\biggl[
\biggl(\frac{2k_{\perp}}{m}+\frac{\alpha k_{\perp}}{k}\biggr)^{2}
+\frac{2g'k^{2}_{\parallel}k_{\perp}(\alpha m+2k)}{mk^{4}}B
+\frac{g'^{2}k^{2}_{\parallel}(\alpha m k^{2}_{\parallel}-3\alpha m k^{2}_{\perp}-6kk^{2}_{\perp})}{\alpha m k^{6}}B^{2}
+\mathcal{O}(B^{3})
\biggr]
\frac{\partial f_{0}}{\partial \varepsilon}~. \nn 
\label{sig0}
\end{eqnarray}
\end{widetext}
Here, $\sum_{\lambda=\pm}$ denotes the sum over the two bands. The $B$-independent and $B$-linear terms in $(\sigma_{\parallel}-\sigma_{\perp})$ involve angular structures proportional to $(k^{2}_{\parallel}-k^{2}_{\perp})$ and $k^{2}_{\parallel}k_{\perp}$, which vanish after $\bm{k}$-space integration. Similarly, the $B$-linear term in $\sigma_{\perp}$ vanishes after momentum integration, while the $B$-independent term remains finite and gives the Drude contribution to the longitudinal conductivity, $\sigma^{\rm D}_{xx}$. Thus, the leading Zeeman-induced planar Hall conductivity is quadratic in the applied magnetic field. In the absence of a magnetic field, the transverse conductivity $\sigma^{\rm Z}_{yx}$ vanishes and the longitudinal conductivity $\sigma^{\rm Z}_{xx}$ is governed by the Drude contribution.

\section{Planar BC and OMM}
\label{Theory of planar BC and OMM}
Here, we define the planar BC and OMM that enable the band geometric PHE in Rashba bilayers. The conventional BC and OMM of the $n$-th band are~\cite{xiao2010berry, Chang_Niu_OMM}
\begin{align*}
{\bm \Omega}_{n\kb} &= {\bm \nabla}_{\kb} \times {\bm {\m A}}_{n\kb}, \\
{\bm m}_{n\kb} &= \frac{ie}{2\hbar} \bra{{\bm \nabla}_\kb u_{\kb}^n} \times (\e_{n\kb} - {\m H}_0)\ket{{\bm \nabla}_\kb u_{\kb}^n},
\end{align*}
where ${\bm {\m A}}_{n\bm k} = \bra{u_\kb^n} i{\bm \nabla}_\kb \ket{u^n_\kb}$ is the Berry connection, $\e_{n\kb}$ is the band energy, and $\ket{u_\kb^n}$ is the periodic part of the Bloch wavefunction for a system governed by Hamiltonian ${\m H}_0$. In an ideal 2D material, carrier motion is confined to the plane, so the BC and OMM vectors have only out-of-plane components. In quasi-2D systems with two or more coupled layers, out-of-plane electronic delocalization can generate in-plane BC and OMM components. These planar components are~\cite{koushik_2dphe, rahul_2dpne}
\begin{subequations}\label{eq:planar_BC_OMM} 
\begin{align}
{\bm \Omega}^{\rm Pl}_{n \bm k} &= 2\hbar \sum_{n' \neq n}\frac{\text{Re}(\bm v_{nn^{'}} \times {\bm {\m Z}}_{n'n})}{\varepsilon_{n\kb}- \varepsilon_{n'\kb}}~,\label{Omega_pl} \\
{\bm m}^{\rm Pl}_{n \bm k} &= e\sum_{n' \neq n} \text{Re}(\bm v_{n n'} \times {\bm {\m Z}}_{n'n})~,\label{m_pl}
\end{align}
\end{subequations}
where the superscript `Pl' denotes the in-plane components. Here, $\bm{v}_{nn'}=(1/\hbar)\langle u^n_{\bm k} | {\bm \nabla}_{\bm k}\mathcal{H}_0 | u^{n'}_{\bm k} \rangle$ is the interband velocity matrix element, and ${\bm {\m Z}}_{n'n} = \langle u^{n'}_{\kb} | \hat{\m Z} | u^{n}_{\kb} \rangle \hat{\bm{z}}$ is the matrix element of the out-of-plane position operator. We use $\hat{\mathcal{Z}}=(c/2)\mathrm{diag}(1,-1)\otimes \sigma_0$, where $c$ is the separation between the two 2DEG layers. Figure \ref{bc_conductivity_plot}(a) shows the planar BC and OMM for equal effective masses and Rashba parameters differing by a factor of two, with $\alpha_{A}=2\alpha_{B}=0.66$~eV$\text{\AA}$.
%%%%%%%%%%%%%%%

\section{Calculation of BC- and OMM-induced PHE current density}
\label{calculation of BC and OMM induced response}

%%%%%%%%%%%%%%%%%%%% Table.1 : Highlighting our 2DPNE responses %%%%%%%%%%%%%%%%%%%%%%%%%%%%%%
\begin{table*}[t!]
\caption{General expressions for the third- ($\chi_{ab;c}$) and fourth-rank ($\chi_{ab;cd}$) BC-OMM-induced planar Hall response tensors. The response tensors are separated into BC, OMM, and mixed (BC + OMM) contributions using Eq. \eqref{current_final}. We write these contributions as $\chi^{\rm BC/OMM/BC+OMM}_{ab;c(d)}  = \frac{e^2}{\hbar} \sum_n \int [d\kb] (A_1 f'_0 + A_2 f''_0 + A_3 f'''_0) + (c \leftrightarrow d)$, where $f'_0$, $f''_0$, and $f'''_0$ are derivatives of the Fermi-Dirac distribution with respect to energy. The term $(c \leftrightarrow d)$ symmetrizes $\chi_{ab;cd}$ in the magnetic field indices. We define $\Omega_{V} = \frac{e}{\hbar}(\bm v_{\bm{k}} \cdot \bm \Omega^{\rm Pl}_{\bm{k}})$ and $\int[d\kb] = \int d^2 k/(2\pi)^2$.}
\centering
\begingroup
\renewcommand{\arraystretch}{1.3}
\setlength{\tabcolsep}{3pt}
\footnotesize
\resizebox{\textwidth}{!}{%
\begin{tabular}{@{}l c l l@{}}
\hline\hline
\multicolumn{4}{c}{Third- and fourth-rank planar Hall response tensors} \\
\multicolumn{2}{c}{Band geometric origin} & $\chi_{ab;c}$ & $\chi_{ab;cd}$ \\
     \hline 
     \multirow{3}{*}{BC} & $A_1$ &
     \(e v_{a}v_{b}\Omega^{\rm Pl}_{c}-\hbar(v_{a}\delta_{bc}+v_{b}\delta_{ac})\Omega_{V}\) &
     \(\begin{aligned}[t]
     -\frac{1}{2\hbar}\bigl(&\hbar^{2}\Omega^{2}_{V}\delta_{ac}\delta_{bd}
     -e\hbar(v_{a}\delta_{bc}+v_{b}\delta_{ac})\Omega^{\rm Pl}_{d}\Omega_{V}\\
     &+e^{2}v_{a}v_{b}\Omega^{\rm Pl}_{c}\Omega^{\rm Pl}_{d}\bigr)
     \end{aligned}\) \\
     & $A_2$ & 0 & 0 \\
     & $A_3$ & 0 & 0 \\
     \hline 
     \multirow{3}{*}{OMM} & $A_1$ &
     \((v_{a}\partial_{b}m^{\rm Pl}_{c}+v_{b}\partial_{a}m^{\rm Pl}_{c})\) &
     \(-\frac{1}{2\hbar}(\partial_{a}m^{\rm Pl}_{c})(\partial_{b}m^{\rm Pl}_{d})\) \\
     & $A_2$ &
     \(\hbar v_{a}v_{b}m^{\rm Pl}_{c}\) &
     \(-\frac{1}{2}(v_{a}\partial_{b}m^{\rm Pl}_{c}+v_{b}\partial_{a}m^{\rm Pl}_{c})m^{\rm Pl}_{d}\) \\
     & $A_3$ & 0 &
     \(-\frac{\hbar}{4}v_{a}v_{b}m^{\rm Pl}_{c}m^{\rm Pl}_{d}\) \\
     \hline 
     \multirow{3}{*}{\begin{tabular}{@{}l@{}}BC +\\ OMM\end{tabular}} & $A_1$ & 0 &
     \(\begin{aligned}[t]
     -\frac{1}{2\hbar}\bigl[&
     e(v_{a}\partial_{b}m^{\rm Pl}_{c}+v_{b}\partial_{a}m^{\rm Pl}_{c})\Omega^{\rm Pl}_{d}\\
     &-e(v_{a}\delta_{bc}+v_{b}\delta_{ac})(\bm{\Omega}^{\rm Pl}\cdot\bm{\nabla}_{\bm k}m^{\rm Pl}_{d})\\
     &-\hbar(\partial_{a}m^{\rm Pl}_{c}\delta_{bd}+\partial_{b}m^{\rm Pl}_{c}\delta_{ad})\Omega_{V}\bigr]
     \end{aligned}\) \\
     & $A_2$ & 0 &
     \(\frac{1}{2}\left[\hbar(v_{a}\delta_{bc}+v_{b}\delta_{ac})\Omega_{V}-e v_{a}v_{b}\Omega^{\rm Pl}_{c}\right]m^{\rm Pl}_{d}\) \\
     & $A_3$ & 0 & 0 \\
     \hline 
     \hline
\end{tabular}
}
\endgroup
\label{table2}    
\end{table*}
%%%%%%%%%%%%%%%%%%%%%%%%%%%%%%%%%%%%%%%%%%%%%%%

In this Appendix, we derive the planar BC- and OMM-driven PHE currents in Rashba spin-orbit-coupled 2DEGs. We start from the semiclassical equations of motion for Bloch electrons in external electric and magnetic fields. The first-order gradient correction to the wave packet gives the BC contribution to the velocity, reflecting the geometric phase of Bloch states~\cite{sundaram_niu_wavepacket}. The self-rotation of the wave packet about its center of mass gives the OMM~\cite{Chang_Niu_OMM}. In an external magnetic field, the OMM modifies the band energy. The BC- and OMM-modified equations of motion are~\cite{xiao2010berry,das2019berry}
\begin{equation}
\label{eq:rdot}
\dot{\bm{r}}_{n}
=
D_{\bm{k}}
\left[
\tilde{\bm{v}}_{n\bm{k}}
+
\frac{e}{\hbar}(\bm{E}\times\bm{\Omega}_{n\bm{k}})
+
\frac{e}{\hbar}\bm{B}
(\tilde{\bm{v}}_{n\bm{k}}\cdot\bm{\Omega}_{n\bm{k}})
\right],
\end{equation}
\begin{equation}
\label{eq:kdot}
\hbar \dot{\bm{k}}_{n}
=
D_{\bm{k}}
\left[
-e\bm{E}
-e(\tilde{\bm{v}}_{n\bm{k}}\times\bm{B})
-\frac{e^{2}}{\hbar}\bm{\Omega}_{n\bm{k}}(\bm{E}\cdot\bm{B})
\right]~,
\end{equation}
where $\tilde{\varepsilon}_{n\bm{k}}=\varepsilon_{n\bm{k}}+\varepsilon^{m}_{n\bm{k}}$ is the OMM-modified $n$-th band energy and $\varepsilon^{m}_{n\bm{k}}=-\bm{m}_{n\bm{k}}\cdot\bm{B}$ is the linear energy correction from the OMM. The OMM-modified band velocity is
\begin{equation}
\tilde{\bm{v}}_{n\bm{k}}
=
\frac{1}{\hbar}\bm{\nabla}_{\bm{k}}\tilde{\varepsilon}_{n\bm{k}}
=
\bm{v}_{n\bm{k}}
+\bm{v}^{m}_{n\bm{k}},
\qquad
\bm{v}^{m}_{n\bm{k}}
=
-\frac{1}{\hbar}
\bm{\nabla}_{\bm{k}}(\bm{m}_{n\bm{k}}\cdot\bm{B})~.
\end{equation}
The phase-space factor is
\begin{equation}
D_{\bm{k}}
=
\left[
1+\frac{e}{\hbar}
(\bm{B}\cdot\bm{\Omega}_{n\bm{k}})
\right]^{-1}.
\end{equation}
Below, we suppress the band index when no confusion can arise.

The occupation function follows from the BTE in the RTA,
\begin{equation}
\frac{\partial f}{\partial t}
+
\dot{\bm{r}}_n\cdot\bm{\nabla}_{\bm r} f
+
\dot{\bm{k}}_n\cdot\bm{\nabla}_{\bm k}f
=
-\frac{f-\tilde{f}_0}{\tau}~,
\label{Boltzmann transport eq}
\end{equation}
where $f=f({\bm r},\bm{k})$ is the non-equilibrium distribution function, $\tau$ is a constant relaxation time, and $\tilde{f}_{0}=[1+e^{\beta(\tilde{\varepsilon}_{\bm{k}}-\mu)}]^{-1}$ is the equilibrium Fermi-Dirac distribution evaluated with the OMM-modified band energy. To linear order in the applied electric field, the distribution function is
\begin{equation}
    f
    =
    \tilde{f}_{0}
    +
    e\tau D_{\bm{k}}\,
    \bm{E}\cdot
    \left(
    \tilde{\bm{v}}_{\bm{k}}
    +
    \frac{e\bm{B}(\tilde{\bm{v}}_{\bm{k}}\cdot\bm{\Omega}_{\bm{k}})}{\hbar}
    \right)
    \frac{\partial \tilde{f}_{0}}{\partial \tilde{\varepsilon}_{\bm k}}~.
\end{equation}

The generalized charge current density is
\begin{equation}
\bm{j}
=
-e\sum_{n}\int [d\bm{k}]
D_{\bm{k}}^{-1}\dot{\bm{r}}_{n}f~.
\end{equation}
Keeping terms linear in $\bm E$, the planar Hall current is
\begin{widetext}
\begin{eqnarray}
\label{current_final}
\bm{j}^{\rm PHE}
&=&
-e^{2}\tau
\sum_{n}
\int[d\bm{k}]\,
D_{\bm{k}}
\tilde{\bm{v}}_{\bm{k}}
(\tilde{\bm{v}}_{\bm{k}}\cdot\bm{E})
\partial_{\tilde{\varepsilon}}\tilde{f}_{0}
-\frac{e^{3}\tau}{\hbar}
\sum_{n}
\int[d\bm{k}]\,
\bm{B}D_{\bm{k}}
(\tilde{\bm{v}}_{\bm{k}}\cdot\bm{E})
(\tilde{\bm{v}}_{\bm{k}}\cdot\bm{\Omega}_{\bm{k}})
\partial_{\tilde{\varepsilon}}\tilde{f}_{0}
\nn\\
&&
-\frac{e^{3}\tau}{\hbar}
\sum_{n}
\int[d\bm{k}]\,
(\bm{E}\cdot\bm{B})
D_{\bm{k}}
\tilde{\bm{v}}_{\bm{k}}
(\tilde{\bm{v}}_{\bm{k}}\cdot\bm{\Omega}_{\bm{k}})
\partial_{\tilde{\varepsilon}}\tilde{f}_{0}
-\frac{e^{4}\tau}{\hbar^{2}}
\sum_{n}
\int[d\bm{k}]\,
(\bm{E}\cdot\bm{B})\bm{B}
D_{\bm{k}}
(\tilde{\bm{v}}_{\bm{k}}\cdot\bm{\Omega}_{\bm{k}})^{2}
\partial_{\tilde{\varepsilon}}\tilde{f}_{0}~.
\end{eqnarray}
\end{widetext}
After expansion in powers of the magnetic field, the planar current density contains first-, second-, and third-order energy derivatives of the Fermi function. It is therefore a Fermi-surface response. For a weak magnetic field, we expand the OMM-modified equilibrium distribution in the energy correction $\varepsilon^{m}_{\bm k}$:
\begin{equation}
    \tilde{f}_0
    =
    f_0
    +
    \varepsilon^m_{\bm k}
    \frac{\partial f_0}{\partial \varepsilon_{\bm k}}
    +
    \frac{1}{2}
    (\varepsilon^m_{\bm k})^{2}
    \frac{\partial^2 f_0}{\partial \varepsilon^{2}_{\bm k}}
    +\cdots~.
\end{equation}
We also expand $D_{\bm{k}}$ as
\begin{equation}
D_{\bm{k}}
\simeq
1-\Omega_{B}+(\Omega_{B})^2+\mathcal{O}(B^3),
\qquad
\Omega_{B}
=
\frac{e}{\hbar}
(\bm{B}\cdot\bm{\Omega}_{\bm{k}}).
\end{equation}
These expansions give the planar current density to linear and quadratic orders in $\bm{B}$. With only conventional out-of-plane BC and OMM components, $\bm{B}\cdot\bm{\Omega}_{\bm k}=0$, $\tilde{\bm{v}}_{\bm{k}}\cdot\bm{\Omega}_{\bm k}=0$, and $\bm{m}_{\bm k}\cdot\bm{B}=0$ for an in-plane magnetic field. All BC- and OMM-induced planar current components therefore vanish in a strictly 2D system. In a quasi-2D bilayer, finite in-plane BC and OMM make $\bm{B}\cdot\bm{\Omega}^{\rm Pl}_{\bm{k}}$, $\tilde{\bm{v}}_{\bm{k}}\cdot\bm{\Omega}^{\rm Pl}_{\bm k}$, and $\bm{m}^{\rm Pl}_{\bm k}\cdot\bm{B}$ nonzero, enabling finite band geometric planar Hall currents. We express the total planar current density in terms of the third- ($\chi_{ab;c}$) and fourth-rank ($\chi_{ab;cd}$) response tensors in Eq.~\eqref{planar_current_expression}. The BC- and OMM-derived $B$-linear and $B$-quadratic response tensors are summarized in Table~\ref{table2}.

\section{Relative size of the two PHE channels}
\label{relative_scale_estimate}
Here, we isolate the factors that dictate the relative size of the Zeeman and band geometric PHE channels. For the symmetry setting considered here, the $B$-linear transverse response vanishes. The leading transverse conductivities can therefore be written as
\begin{align}
    \sigma^{\rm Z}_{yx}(\mu,\theta) &= \tau B^{2}\Lambda_{\rm Z}(\mu)\sin{\theta}\cos{\theta}~, \label{sigma_z_scale}\\
    \sigma^{\rm BG}_{yx}(\mu,\theta) &= \tau B^{2}\chi_{yx;xy}(\mu)\sin{\theta}\cos{\theta}~. \label{sigma_b_scale}
\end{align}
Here, $\Lambda_{\rm Z}(\mu)$ is the coefficient of the $B^{2}$ part of $(\sigma_{\parallel}-\sigma_{\perp})/\tau$ generated by the Zeeman-induced velocity anisotropy, while $\chi_{yx;xy}(\mu)$ is the band geometric tensor in Eq. \eqref{planar_current_expression}. Away from angles where $\sin{\theta}\cos{\theta}=0$, the ratio of the two responses is
\begin{equation}
    {R}(\mu)\equiv
    \left|\frac{\sigma^{\rm Z}_{yx}}{\sigma^{\rm BG}_{yx}}\right|
    =
    \left|\frac{\Lambda_{\rm Z}(\mu)}{\chi_{yx;xy}(\mu)}\right|~.
    \label{relative_ratio}
\end{equation}
Thus, within the weak-field and constant-$\tau$ approximation used here, the useful diagnostic is ${R}(\mu)$. The common factors $\tau$, $B^{2}$, and $\sin{\theta}\cos{\theta}$ do not distinguish the two mechanisms.

The Zeeman prefactor is controlled by the field-induced distortion of the Rashba Fermi surface. A useful local estimate is given by the ratio of Zeeman energy to Rashba spin-splitting energy scale
\begin{equation}
    \eta_{\rm Z}(k_F)=\frac{\varepsilon_{\rm Z}}{\varepsilon_{\rm R}(k_F)}
    =\frac{g'B}{\alpha k_F}~,
\end{equation}
where $\varepsilon_{\rm Z}=g'B$ and $\varepsilon_{\rm R}(k_F)=\alpha k_F$ is the Rashba spin-orbit energy at the Fermi wave vector. 

The band geometric prefactor is controlled by the planar BC and OMM. From Eqs.~\eqref{Omega_pl} and \eqref{m_pl},
\begin{equation}
    \Omega^{\rm Pl}_{n\bm k}\sim
    \hbar \frac{v_{nn'} {\cal Z}_{n'n}}{\Delta_{nn'}(\bm k)}~, \qquad
    m^{\rm Pl}_{n\bm k}\sim e v_{nn'} {\cal Z}_{n'n}~,
\end{equation}
where $\Delta_{nn'}(\bm k)=\e_{n\bm k}-\e_{n'\bm k}$ is the relevant interband separation. The BC-OMM channel is enhanced when interlayer delocalization gives finite ${\cal Z}_{n'n}$ matrix elements and Rashba asymmetry produces sizable planar BC and OMM near small interband gaps. It is suppressed when the layers are effectively decoupled, or when the relevant interband gaps are too large.

This analysis indicates that the Zeeman response dominates when $|\Lambda_{\rm Z}|\gg |\chi_{yx;xy}|$, while the band geometric response becomes competitive when interlayer delocalization, Rashba asymmetry, and small interband gaps make $|\chi_{yx;xy}|$ comparable to $|\Lambda_{\rm Z}|$.

\bibliography{reference.bib}
\end{document}